\documentclass[10pt, letterpaper]{article}
\usepackage{amsfonts, mathrsfs}
\usepackage{graphicx, color}
\usepackage[dvips]{epsfig}
\newcommand{\apj}{ApJ}
\newcommand{\apjl}{ApJ}
\newcommand{\aap}{A\&A}
\newcommand{\jcoph}{J. Comp. Phys.}

\newcommand{\p}[1]{(\ref{#1})}

\def\2s{{$N=2$ SUSY\ }}

\def\ma{\mbox{$\mathcal{A}$}}

\def\l2{{\mbox{$SL(2,{\mathbb Z})$}}}

\def\l{ \Lambda}

\def\k{ \mathrm{K}}
\def\a{ \mathrm{Ai}}
\def\b{ \mathrm{Bi}}
\def\fb{ \bar{f}}
\def\bd{ \bar{d}}
\def\bl{ {\bar{\lambda}} }
\def\tl{\tilde{\Lambda}}

\begin{document}

%\title{Flame-capturing technique. 1: Adaptation to gas expansion}
%\author{Andrey V. Zhiglo}
%\date{29 Aug 05}
%\maketitle
\begin{center}
{\Large Flame-capturing technique. 1: Adaptation to gas expansion}\\ \bigskip
{\large Andrey V. Zhiglo}\\ \medskip
Center for Astrophysical Thermonuclear Flashes and Department of Physics, University of Chicago\\
5640 S. Ellis Ave., Chicago, IL 60637, USA; {\tt azhiglo@uchicago.edu}
\end{center}
\begin{abstract}
Various flame tracking techniques are often used in hydrodynamic
simulations. Their use is indispensable when resolving actual scale of
the flame is impossible. We show that
parameters defining \lq\lq artificial flame\rq\rq\ propagation found
from model systems may yield flame velocities several times distinct
from the required ones, due to matter expansion being ignored in the models.

Integral effect
of material expansion due to burning is incorporated into flame capturing
technique (FCT) \cite{X95}. 
Interpolation formula is proposed for the parameters governing flame
propagation yielding 0.2\% accurate speed and width for any expansion 
(and at least 0.01\% accurate for expansions typical in type Ia supernova explosions.) 
Several models with simple
burning rates are studied with gas expansion included. Plausible performance
of the technique in simulations is discussed. Its modification ensuring
finite flame width is found.
Implementation suggestions are summarized, main criterion being the scheme
performance being insensitive to expansion parameter (thus absence of systematic
errors when the burning progresses from inner to outer layers); in this
direction promising realizations are found, leading to flame structure
not changing while flame evolves through the whole range of densities in the
white dwarf. 
\end{abstract}

\section{Introduction}

In modeling astrophysical flames one is often faced with the necessity
to properly track discontinuity surfaces. A classical situation is
simulating deflagration in a near Chandrasekhar-mass white dwarf (WD),
in currently a \lq\lq standard model\rq\rq\ of Ia-type supernova (SN Ia)
phenomenon. Typical width of a flame\footnote{i.e. a region
where most energy is released, and where chemical composition mostly
changes from \lq\lq fuel\rq\rq\ to \lq\lq ash\rq\rq\ of a thermonuclear
reaction of interest; usually one considers C+O burning to intermediate
mass elements or, at lower densities (when quasiequilibrium timescale
is long compared to flame passing time) to Mg, Ne, He and H.
The \lq\lq ash\rq\rq\ may undergo further metamorphoses after the flame passes,
effect of that on the flame structure and speed w.r.t. fuel neglected.}
is $< 1$mm \cite{TimWoo92}, which cannot be resolved, as a total size of
the box for hydrodynamic simulations is of the star scale, $>10^8$cm.

To get energetics right one needs somehow recreate flame geometry
evolution without resolving responsible for the latter microscales.
Such a problem is not untypical in physics; one hopes to mimic the
processes without a cutoff (= crude grid cells, of size $\Delta$) by
cleverly prescribing velocity of the flame $D_f(\Delta)$ and perhaps other
parameters, making them cutoff-dependent, so that the resulting evolution
coincided with (discretized) evolution of the exact system. At the current
stage several prescriptions for $D_f(\Delta)$ are in use. They are based
on simple physical models of hydrodynamic instability development
(essentially a Rayleigh-Taylor instability modified by burning; see
\cite{X94} and \cite{X95}, where the proposed $D_f(\Delta)$ is also
verified numerically by modeling turbulent burning in uniform gravity and
grid resolving instability scale), yielding $D_f$ of order of turbulent
velocities on scale $\Delta$,
$S_\mathrm{t}\simeq 0.5\sqrt{\ma g \Delta}$, \ma\ denoting the Atwood
number for fuel--ash, $g$ being the gravitational acceleration (or its
projection onto the flame normal).
For more general recipes for $D_f$ for an arbitrary grid (giving essentially
the same values for uniform Cartesian grids) see \cite{clement, NiHi95};
\lq\lq renormalization invariance\rq\rq\ of these prescriptions was
numerically checked in \cite{Rein02}, though few details are provided. 
Global performance of the numerical integration scheme at different
resolutions is presented in detail in \cite{X03,X05}. With $D_f(\Delta)$ defined
as a maximum of the above $S_\mathrm{t}(\Delta)$ and laminar flame speed these
simulations (3D with octahedral symmetry) show that the energy released
at $\Delta_{\mathrm{min}}=5.2\times 10^5\:$cm (adaptive grid used) differs
from that at $\Delta_{\mathrm{min}}=10.4\times 10^5\:$cm by $<8$\% within the
first 2 seconds after ignition (and is practically indistinguishable before
$t=1.62\:$s, the moment when detonation was triggered). Perhaps, more
direct checks would be desirable (this is particularly important for showing
the validity of adaptive-grid simulations), but it is questionable if the
program outlined in the beginning of the paragraph is workable at all in
view of intrinsic non-stationarity of the problem, instabilities and
numerical noise. The cited works provide evidence of this invariance in
statistical sense, which seems adequate in this kind of problems. 
All these treatments share
(in 3D, on scales $\Delta > 10^5$ cm) a remarkable property of
actual $D_f(\Delta)$ being insensitive to microphysics, with no direct dependence
on laminar flame speed.\footnote{
(See \cite{X95} for physical discussion of self-regulation mechanism
behind this phenomenon.) This, in fact, leads to low sensitivity of
mass-burning rate to the
$D_f(\Delta)$ used as long as resolution is enough to get 2--3 generations
of bubbles formed before substantial expansion. That was really observed in
\cite{X03}: multiplying $S_\mathrm{t}$ by $\sqrt{2}$ increased released nuclear
energy, kinetic energy and burned fraction by less than $3\%$ for the
highest resolution used ($\Delta x_\mathrm{min}=2.6\:$km), but for
$\Delta x_\mathrm{min}=10.4\:$km the effect was much more pronounced, about 30\%,
and the difference increased in time up to $1.8$ sec showing that self-regulation
had not been achieved.}

After having chosen the favorite prescription for $D_f(\Delta)$ one starts
evolving the flame surface in time; basically, each element of the flame
is advected by the local material velocity field plus propagates with
$D_f$ normal to itself into the fuel. One approach (\emph{level set method}, LSM)
is to represent a flame
surface as the zero-level of a certain function $G$: flame
manifold$|_t=\lbrace\mathbf{r}(t):G(\mathbf{r},t)=0\rbrace$; one then
prescribes time evolution for $G$ ensuring the needed evolution of
the flame surface plus some additional physical conditions, usually
taking care of $G$ not developing peculiarities hindering computations.
Thus at any time the flame is infinitely thin, grid cells it intersects
contain partly fuel, partly ash. While such a picture seems
physically-motivated, the exact fractions of fuel/ash concentrations,
fluxes, etc. are computed rather crudely, the PPM-based hydrodynamic
solvers used \cite{cowo} spread discontinuities over several grid spacings
anyway. In the present paper I will stick to another flame-tracking
prescription \cite{X94}.\footnote{see \cite{vn} for the original
realization.}

This \emph{flame-capturing technique} (FCT) is a \lq\lq physical\rq\rq\
way of evolving the flame\footnote{various
simulations confirm numerically basic equivalence of the two methods \cite{Rein02,
 Reinlanl02}. In specific situations one method may have advantages over
the other one. Just \lq\lq by construction\rq\rq\ no additional steps must
be undertaken to retain sensible \lq\lq flame\rq\rq\ width in
FCT; in view of the discussion in the beginning one may also hope that this
\lq\lq flame\rq\rq\ response to hydrodynamic factors will reasonably
match that of the real turbulent burning region, if implemented properly.}. 
On
increasing dissipative terms (and, actually, reintroducing them, as
the only relevant in degenerate fuel burning thermal conductivity
term is negligible any far from the discontinuity) in hydrodynamic
equations one broadens the flame. Hence the strategy to choose such
terms so that to make the flame width equal to a few grid spacings
while the flame propagation velocity having the desired value. One
can then smoothly turn the dissipative terms off far from the flame.
In \cite{X95} this idea was implemented by adding an equation for
conventional \lq\lq fuel\rq\rq\ concentration (denoted
$1-f(\mathbf{r})$ there) to basic hydrodynamic set of
equations (continuity with known sources for the species concentrations,
momentum and energy),
\begin{equation}\label{I1}
    \frac{\partial f}{\partial t}+\mathbf{u}\cdot\nabla f
    =K\nabla^2f+\Phi(f)
\end{equation}
with the simplest second space-coordinate derivative diffusion-like term
and step-function burning rate
$\Phi(f)\sim\theta(f-f_0)(1-\theta(f-1))$;\footnote{we define
$\theta(x)=1$ if $x\ge 0$, $\theta(x)=0$ if $x<0$.} $\,f\in[0;1]$,
$f=0$ corresponding to pure \lq\lq fuel\rq\rq; $\textbf{u}$ is a local
gas velocity. The source term in the energy conservation equation
then read $q\,d f/d t$, $q$ being total heat production
under burning of the fuel (C+O). $K$ and $\Phi$ were chosen
as described above: dependence of the velocity of the
flame \p{I1} on $K$ and $\Phi$ was obtained by analytically solving
\p{I1} in 1 dimension, with $\mathbf{u}$ being constant, simply related
to the $D_f$; flame width was in effect estimated crudely to be of order
$\sqrt{K/R}$, its dependence on $f_0$ was not studied. This scheme
has been used essentially verbatim by several groups since then.

Apart from an insignificant (at the value $f_0=0.3$ used) error
in determining $D_f(K,\Phi)$ in \cite{X95} there are good
reasons to study system \p{I1} further:
\begin{itemize}
    \item As described above, the technique is quite crude. The
    prescription for $D_f$ must vary along the flame front. As applied
    before (see thorough description in \cite{X00}), the only factors 
    influencing the $D_f$ were local gravity and mutual orientation of the front
    element and
    gravitational field. In reality the flame structure/speed are
    influenced by other factors as well, even neglecting the role of
    thermodynamic parameters variation along the front\footnote{assuming
    $\sqrt{\ma}$ in the $S_\mathrm{t}$ definition taking care of the bulk of that
    dependence.}: curvature of the front; independent
    of the curvature flame stretch, resulting from shear velocity
    gradients. Aside from these, even in 1D the above procedure has
    flaws. Gas velocity $\mathbf{u}$ in \p{I1} will change significantly
    \emph{within} the flame due to the gas expansion as a result of
    burning, this will make the actual flame propagation speed less than
    the designed $D_f$, based on which $K$ and $\Phi$ were constructed as
    described above.
    We suggest that the above factors affect the $D_f$ in some universal
    way, independent of the underlying
    microstructure; that is model system \p{I1} must be affected
    in much the same way as intricately wrinkled thin flame provided
    the widths and velocities of resulting burning regions coincide
    on scale $\Delta$.\footnote{
  of course, relations like scaling $D_f^2\propto\,$flame width will not
  hold for \p{I1}-like systems as the former are critically dependent
  on the mechanism behind the structure of turbulence in gravitational
  field. Yet the hope is that the role of gravity is just to set scales
  of turbulent motions and hence the effective transport coefficients
  (like eddy viscosity/diffusivity), and knowing those alone is enough to
  model the effects concerned in the main text.}

    Ultimately this study would provide one with the prescription how to modify
    the parameters in \p{I1} based on local density, gravity, flame curvature,
    background velocity fields, etc.  so that to match the response of the flame 
    region to major hydrodynamic factors in real life. 
    
    \item Apart from improving the FCT
    reliability this study will show the amplitude/relevance of the
    mentioned effects in flame systems, a matter of constant
    attention \cite{markstein,Dursi03}, and how different is the effect of
    these on different test systems.
\end{itemize}

In this paper I study the effect of gas expansion on different
realizations of system \p{I1} in 1D. The higher-dimensionality effects
require completely different methods, the results will be reported
separately. In Section 2 the problem is formulated, the numerical
procedure is described. 
In Section 3 we qualitatively analyze the master equation \p{I1} of
FCT with matter expansion for parabolic (KPP) and step-function $\Phi(f)$;
difference in flame profiles helps to formulate criteria for selecting a
viable realization of \p{I1} promising adequate performance in non-stationary
simulations. 
Section 4 contains analytic investigation of the problem for $\Phi$ being
step-function, confirms accuracy of the numerical procedure and provides
much faster way of getting the parameters; a new form of the diffusion term
is proposed, ensuring finite flame width.
Section 5 includes simple astronomically-accurate fits for the data obtained
in Sec.~4 valid for any expansion; these are then refined to the SN Ia problem,
we summarize a few \lq\lq best\rq\rq\ (in our opinion) schemes, explicit
prescriptions for coding are provided.
We conclude in Sec.~6. Appendix contains a simple analysis of performance
of the scheme in real simulations.

This work expands an earlier investigation by the author presented at FLASH
Center site review (University of Chicago, Oct. 2004; unpublished), comments
on the performance in non-steady simulations and approximate formulae for
parameter values are added. Different from \cite{X95} form of the
diffusion term (first term on the R.H.S. of \p{I1}) is also studied
here for the first time.

\section{Formulation and the numerical method}
\subsection{Boundary problem for finding $D_f[\Phi(f);\Lambda]$}
In this paper we restrict ourselves to a simple one dimensional model,
the one where all the quantities can be expressed algebraically as
functions of the fuel concentration field only\footnote{
this applies,
say, for similar concentration and temperature fields.
%It is easy to show
%(say, by trivially modifying an argument in \cite{bychkov95}; this is a
%manifestation of the general property observed in \cite{z-fk}) that often
%used approximation when only one reaction proceeds in the front falls into
%this category.
It is known \cite{z-fk} that often used approximation when only one reaction
proceeds in the front falls into this category.

$\:$ Remarks on physical nature of the constructions presented are made in
several places in this paper. These are more than plain motivation; though the
main effect -- dependence of the artificial
flame speed on expansion as a result of burning can be correctly taken into
account with many bizarre variations of \p{I1} (all of these yielding the
needed flame speed in steady-state problems, with further caveats like
matching real $\mathbf{u}$ in \p{I1} with its models, say
(\ref{front}--\ref{rho1})) the toy systems properly modeling
averaged distribution of density and concentration of the principal reactant (or,
usually equivalently, distribution of heat release) in real flames
can fairly be expected to respond in the same way to the non-stationary factors
ignored in this paper. We will be able to estimate only some of the latter in the
further publications, so at any stage the naturalness of the model will remain
the argument that the effects ignored on physical grounds are not amplified in the unphysical model beyond what physical intuition could suggest.}.
The pressure is considered hydrostatic, as in deflagrations matter
velocities are at least an order of magnitude smaller than the sound speed,
apart maybe from the outer regions of the WD \cite{TimWoo92}. The diffusion
equation we consider reads
\begin{equation}\label{artvis0}
\frac{\partial(\rho f)}{\partial t}+\frac{\partial(\rho fu)}{\partial x}= \rho
K\frac{\partial ^2f}{\partial x^2}+\rho\Phi(f),
\end{equation}
$f$ being the concentration of the reaction product ($1-f$ that of the fuel), $u$ local velocity,
$\rho$ density of the mixture and $K$ the diffusivity.

We are searching for self-similar solutions (all quantities
depending on $x-D_ft$ only); as applied in FCT this means we neglect
thermodynamical variables changing (w.r.t. flame position) on
timescales of establishing such a steady-state burning regime. In the
system of reference where the flame rests the gas velocity $u(x)$
monotonically changes from $-u_0-D_f$ as $x\to-\infty$ to $-D_f$ at
$x=+\infty$, most of this change happening within the flame width;
mass conservation allows to express $u$ 
\begin{equation}\label{front}
u(x)\rho(x)=-D_f\rho_0,
\end{equation}
($\rho_0$ being the density of the fuel far ahead of the front.)
Substituting the self-similar form of solution and the $\rho$
\begin{equation}\label{rho1}
\frac{\rho_0}{\rho}=1+\frac{\gamma-1}\gamma\,\frac{q\rho_0}{P_0} f
\end{equation}
(enthalpy conservation; ideal gas EOS used in this section, $q$ signifies the heat produced
in the reaction) into \p{artvis0} yields a nonlinear equation for $f$:
\begin{equation}\label{artvisfront0}
-D_f\frac{\rho_0}{\rho}\frac{df}{dx}=K\frac{d^2f}{dx^2}+\Phi(f);
\end{equation}
$\Phi$ throughout stands for the burning rate. We mainly consider two
model $\Phi$: \cite{X95}
\begin{equation}\label{Phistep}
\Phi=\left\lbrace
\begin{array}{ll}
  R &\: \mathrm{for\ } f_0<f<1\\
  0 &\: \mathrm{otherwise}
\end{array} \right.
\end{equation}
and \cite{KPP}
\begin{equation}\label{Phikolm}
\Phi=R\, f(1-f)\quad (\mathrm{for\ }0<f<1).
\end{equation}
The goal is to find eigenvalues of $D_f$ for the boundary problem,
\p{artvisfront0} (rewritten using \p{rho1} as
\begin{equation}\label{masterKR}
\frac d{d x}\left\lbrace K\frac{df}{dx}+D_f\left( f+\frac{f^2}{2\Lambda}\right)
\right\rbrace=-\Phi(f),
\end{equation}
for analytic consideration) with requirements\footnote{$f$ must be
continuously differentiable; left and right limits of $d^2f/dx^2$
will be different wherever $f=f_0$ for the step-function $\Phi$.}
\begin{eqnarray}
f(-\infty)&=&1 \label{bndl} \\
f(+\infty)&=&0 \label{bndr}\\
0\le f\le 1 &&\forall x\in\mathbb{R}.\label{bndm}
\end{eqnarray}
For applications in the FCT $\Lambda$ in \p{masterKR} must be chosen based on
the integral density jump across the flame,%
\footnote{$\Lambda=\frac\gamma{\gamma-1}\,\frac{P_0}{q\rho_0}$ would hold
were the fuel ideal gas. Even though for generic fuel $\frac{\rho_0}{\rho}(f)$
is nonlinear we expect the $D_f$ based on $\frac{\rho_0}{\rho}=1+\frac f \Lambda$
with \p{lamb} accurate enough; see Appendix for more.}
\begin{equation}\label{lamb}
\Lambda=\frac{\rho_\mathrm{ash}}{\rho_\mathrm{fuel}-\rho_\mathrm{ash} }
\equiv 0.5(\mathcal{A}^{-1}-1).
\end{equation}
To typical WD densities
$\rho_\mathrm{fuel}\in[3\times 10^7;3\times 10^9]$ $g/cm^3$ correspond
$\Lambda$ from 0.69 to 6.3 (data adapted from \cite{X95}, for 0.5C+0.5O
composition).

Bearing in mind the localized character of heat production/expansion under real
(Arrhenius-type) burning, it may seem logical to improve our linear \p{rho1} like

\begin{equation}\label{rho2}
\frac{\rho_0}{\rho}=1+\frac{f-f_0}{\Lambda(1-f_0)}\,\theta(f-f_0)
\end{equation}
with $\rho$ remaining unchanged in the region $f\le f_0$ without
burning (even though $f$ changes with $x$ due to diffusion); fuel
not expanding in long diffusion tail (see below) really yields $D_f$
significantly less sensitive to $\Lambda$; this is accompanied with
broader flames. Though the latter situation may seem more convenient
there is no clear way to generalize \p{rho2} to more general $\Phi$; moreover,
as shown in the Appendix, \p{rho1} is far better motivated.
Below we study both models (called I and II for short) to get feeling
of the sensitivity of the results to the EOS.

\subsection{Numerical procedure}
Upon separating $\Phi$ into a product of scale- and form-factors,
$\Phi(f)=R\Phi_0(f)$ one may deduce the explicit dependence of $D_f$
(supposedly existing eigenvalues) on $R$ and $K$: namely
\begin{equation}\label{Dfscale}
D_f=d \sqrt{KR} ,
\end{equation}
where $d$ is to be found as an eigenvalue of the boundary problem for $f$
considered a function of a dimensionless coordinate
\[ \chi=x\sqrt{\frac{R}{K}}
\] satisfying the following equation:
\begin{equation}\label{master}
\frac {d^2f}{d \chi^2}+d\left(1+\frac
f\Lambda\right)\frac{df}{d\chi}+\Phi_0(f)=0,
\end{equation}
and boundary conditions like (\ref{bndl}--\ref{bndm}).

As (\ref{master}) is invariant under translations in $\chi$ it can generically
be reduced to a first order equation by rewriting it in terms
of
\[ p=-\frac{df}{d\chi} \] considered a function of $f$:
\begin{equation}\label{master1}
\frac{dp}{df}-d\left(1+\frac f\Lambda\right)+\frac{\Phi_0(f)}{p}=0.
\end{equation}
This form is used below for qualitative and numerical analysis of
the problem. Corresponding boundary conditions are $p(0)=p(1)=0$.
Eigenfunctions $p(f)$ are nonnegative.

To diminish the effect of singularity at $p=0$, \p{master1} was
rewritten in terms of $v=p^2$,
\begin{equation}\label{master12}
\frac 1 2\frac{dv}{df}=d\sqrt{v}(1+f/\Lambda)-\Phi_0(f).
\end{equation}
In order not to pay extra attention to locating $v=0$ points,
$\sqrt{v}$ was replaced with $\sqrt{|v|}\:$\footnote{
 A side effect was instability of the solution next to $f=1$ with
 respect to becoming negative with fast growing absolute value.
 % -- this can be seen after rewriting \p{master12} in terms of
 %$\fb=1-f$.
To account for this $v$ was set to 0 whenever it became
negative near $f=1$. The procedure got stabilized fast, the
resulting integral curves closely followed analytically predicted
behavior, ensuring the reliability of the
algorithm. The analytic asymptotes were used in later realizations
as seed values for $v$ near 0.}.
\p{master12} was integrated by the fourth order Runge-Kutta
algorithm starting from $f=1,p=0$.\footnote{It is imperative to
start from $f=1$ for parabolic $\Phi_0(f)$ as a general solution
for $f(x)$ near $x\to+\infty$ (where $f\to 0$) is a superposition of
two decaying exponents, and the faster decaying one is lost when integrating
$dp/df$, thus making it impossible to satisfy $p|_{f=1}=0$.}
Grid spacing was constant
($\Delta f=10^{-5}$ for most of the runs), except near the
starting point and wherever a
relative change in $v$ as a result of a single iteration exceeded
$1/16$ where it was refined further.

The $d$ eigenvalue was then found {for step-function $\Phi_0(f)$} by
requiring $p|_{f=1}=0$. Namely, Newton-Raphson algorithm (see, e.g.
\cite{numfor}) was applied, $\frac{\partial
\left(\vphantom{p^2}v(f=1)\right)}{\partial d}$ was found
simultaneously with $p(f)$, ensuring fast convergence.
$d(\Lambda=\infty)$ (found beforehand by solving \p{Lambdainf}) was
used as a seed at each new $f_0$ value for the first $\Lambda$ in the
row, for subsequent $\Lambda$\rq s the previous one provided seed value
for $d$; 4--13 iterations were enough to get $d$ with $10^{-8}$
precision.

%%%%%%%%%%%%%%%%%%%%%%%%%%%%%%%%%%%%%%%%%%%%%%%%%%%%%%%%%%%%%%%%%%%%%%%%%%%%%%%
\section{Qualitative behavior at different burning rates}
\subsection{Step-function}
We start with the model described by (\ref{master}) with
$\Phi_0(f)=(1-\theta(f-1))\theta(f-f_0)$. The
self-similar solution describing flame propagation we are
interested in must behave as follows:
%There exists a solution of (\ref{master}) with
\begin{equation}\label{leftb}
f=1 \;\forall
x<x_1,\,\;df/dx\,(x_1)=0,\;0<f<1\mbox{ at }x>x_1,\;\lim\limits_{x\to\infty}f(x)=
0
\end{equation}
(moreover, it is monotonically decreasing, in accord with physical
expectations; see typical flame profiles in Fig.~1). Proof: as any solution
of (\ref{master}) equal to 1 at any point with non-zero $df/d\chi$
would necessarily exceed 1 nearby, the only other possible behavior
near $f=1$ would have been a solution approaching $f=1$
asymptotically as $\chi\to -\infty$.
The behavior of the solution of (\ref{master}) in that region
would have coincided with that of the linearized equation,
\begin{equation}\label{f1step}
f=\bar{c}_1-\chi/\bar{d}+\bar{c}_2\exp(-\bar{d}\chi),
\end{equation}
where
\begin{equation}\label{dbar}
\bar{d}=d\left(1+1/\Lambda\right).
\end{equation}
The latter does not remain in the vicinity of 1 as $\chi\to -\infty$ (in fact is
unbounded) for any values of the integration constants $\bar{c}_{1,2}$; the only
possibility for a solution to stay in $[0;1]$ is to become identically 1 at
$\chi$ smaller than some $\chi_1$, differentiably glued to $f$ behaving like
\p{f1step} as $\chi\to\chi_1+$.

% thus we conclude that \emph{every} solution of our boundary problem is
%identically 1 on some half-line.

Similar analysis in the region where $|f|\ll 1$ yields general solution of
linearized equation
\begin{equation}\label{f0step}
 f=c_1+c_2\exp(-d\chi) ,
\end{equation}
which tends to 0 at $\chi\to -\infty$ iff $c_1=0$.\footnote{Thus any solution of
\p{master} such that $f(\infty)=0$ cannot equal
zero at any finite point. It has an infinite tail,
decaying exponentially.} This completes the proof.%
\footnote{As to $f$ monotonousness, one can observe even more, that $f(x)$ is convex
at $f<f_0$ and concave elsewhere. Really, the solution of \p{master1}
at $f<f_0$ is $p=df(1+f/2\Lambda)$, positive and increasing. If $p$ were not
monotonically decreasing in $[f_0;1]$, one might consider $f_r$, the largest
$f$ s.t. $dp/df|_{f_r}=0$ (it exists as $dp/df$ is continuous in $(f_0;1)$
and negative near 1): $p(f_r)>0$, $d^2p/df^2|_{f_r}=d/\Lambda>0\Rightarrow p(f=1)>0$,
contradicting the boundary condition.}

\begin{figure}[htbp] \begin{centering}%\vspace*{-4cm}\hspace*{-20mm}
\includegraphics[width=0.5\textwidth]{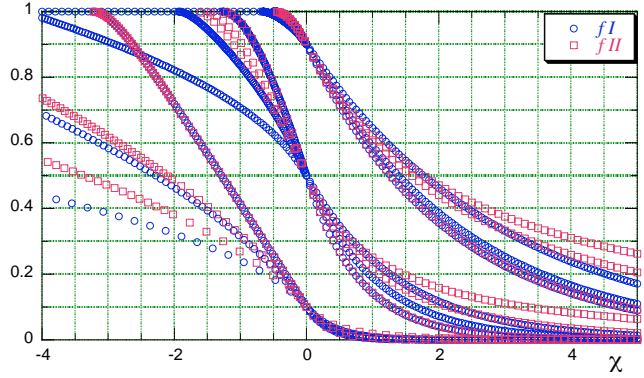}
\caption{Flame profiles ($=$numerically found eigenfunctions). For each $f_0$
(the ordinate of the curve intersection with $\chi=0$) 3 pairs of profiles
are depicted, for $1/\Lambda=0.05,\,4,$ and 20, larger $1/\Lambda$
corresponding to the curves intersecting $\chi=0$ axis at larger
angles. Blue circles represent profiles for model I, red squares
those for model II.}
\end{centering}\end{figure}

Summing up, the desired eigenfunction asymptotically behaves like (\ref{f0step})
with $c_1=0$ as $\chi\to +\infty$, and like (\ref{f1step}) as $\chi\to \chi_1+$
($\chi_1=x_1\sqrt{R/K}$), with $\bar{c}_{1,2}$ such that
\begin{equation}\label{lbdry}
 f(\chi_1)=1,\; df/d\chi(\chi_1)=0.
\end{equation}

Presence of the three restrictions on the two-parameter set of functions
(general solution of (\ref{master})) provides a hint that only under special
circumstances does the required solution exist\footnote{more precisely: for any
fixed $\chi_1$ the solution $f_1$ satisfying (\ref{lbdry}) is unique. Another
boundary condition being vanishing of $f$ at $\chi\to +\infty$ makes the choice
of $\chi_1$ immaterial. That boundary condition is satisfied by one-parameter
subset of solutions, and in order for $f_1$ to belong to that subset generically
there must be one functional dependence among the parameters in \p{master}.}. As
it is shown below, this actually leads to a unique value for $d$ for any fixed
values of the other parameters ($\Lambda$ and $f_0$ here), for which there exist
solutions of the boundary problem.

In the $q=0$ ($\Rightarrow\rho(x)=\rho_0=\mbox{const}$) case 
(\ref{master}) actually \emph{is} piecewise linear, thus
the above three restrictions immediately yield
$d(\Lambda=\infty,f_0)$ as the solution of equation (cf. \p{xi1};
$f(\chi_1)$ is then expressed in elementary functions as $\Lambda=\infty$)
\begin{equation}\label{Lambdainf}
f_0d^2=1-e^{-d^2}.
\end{equation}
One can write the solution in the limit $f_0\to 0$ asymptotically as
$d^2=\frac 1{f_0}-\frac {e^{-1/f_0}}{f_0}-\frac
{e^{-2/f_0}}{f_0^2}-\left(\frac 3{2f^3_0}+\frac 2{f^2_0}+
\frac 2{f_0}\right)e^{- 3/f_0}+O\left(
(f_0 e^{1/f_0})^{-4}\right)$; at $f_0=0.3$ this yields $d$ 2\% smaller
than the value in \cite{X95}; at $\Lambda$\rq s of interest the difference will be
more significant. In the limit $f_0\to 1$ $d$ vanishes
as $d^2=2(1-f_0)+\frac 2 3 (1-f_0)^2+\frac 7 9
(1-f_0)^3+\ldots\;$(leading term agreeing with an estimate in
\cite{zelbook}, p.266).

 With minor changes all said above applies to the model \p{rho2} with
$\rho=\rho_0\;\forall f<f_0$, qualitative conclusions being the same.

%%%%%%%%%%%%%%%%%%%%%%%%%%%%%%%%%%%%%%%%%%%%%%%%%%%%%%%%%%%%%%
\subsection{Burning rate \p{Phikolm}}
It is known (see \cite{zelbook} for detailed discussion) that the spectrum
of eigenvalues of the KPP problem is continuous, comprising all reals
above some $d_1$. In this section we show that the same holds if one
includes the term arising from gas expansion, find similar spectrum for a wide
range of $\Phi$\rq s and verify these conclusions numerically.

One can qualitatively analyze the spectrum for the burning rate \p{Phikolm}
along the same lines as for the \p{Phistep} described in the previous section.
Upon linearization in the region where $1-f\ll 1$ ($\chi\to\infty$)
\p{master} yields
\begin{equation}\label{kleftasym}
f=1-\bar{b}_+e^{-\bar{\lambda}_+\chi}-\bar{b}_-e^{-\bar{\lambda}_-\chi},\quad
\bl_\pm=\frac\bd 2\pm\sqrt{\frac{\bd^2}4+1},
\end{equation}
$\bar{d}=d(1+1/\Lambda)$ as before; thus there is a one parameter subset of physical solutions, those behaving
asymptotically like \p{kleftasym} with $\bar{b}_+=0$.

By linearizing \p{master} in the region where $f\ll 1$ one gets asymptotic
behavior of a solution
\begin{equation}\label{krightasym}
f\approx {b}_+e^{-{\lambda}_+\chi}+{b}_-e^{-{\lambda}_-\chi},\quad
\lambda_\pm=\frac d 2\pm\sqrt{\frac{d^2}4-1}.
\end{equation}\\
There are a priori three different situations for drawing further conclusions:\\
\indent 1) $d<2$: any solution getting to the neighborhood of 0 necessarily
becomes negative. \\
\indent 2) $d>2$: any $b_+,b_-\ge 0$ describe physically acceptable
behavior, as does a certain subset of $b_+<0,\;b_->0$. Thus one may
conjecture that \emph{for all} $d>2$ there is an
eigenfunction: it belongs to the described above 1-parameter subset
of solutions asymptotically approaching 1 as $\chi\to -\infty$
and on becoming small
at larger $\chi$ it behaves asymptotically like \p{krightasym},
exponentially approaching zero as $\chi\to+\infty$. Essentially, this
1-parameter set of solutions corresponds to one of them translated
arbitrarily along $\chi$, i.e. there is a unique flame profile for
any $d$ (the term \lq\lq unique\rq\rq\ is used below in this sense,
i.e. up to translations).

One can show that the solution cannot approach any value in $(0;1)$
as $x\to+\infty$ (say, by linearizing \p{master} near such a value;
this is of course obvious on physical grounds), and by analyzing
\p{master1} one expects the $f$ to decrease monotonically. Thus for
the conjecture to fail, the solution of the above set (\lq\lq
physical\rq\rq\ near 1) must have an asymptote near 0 with unphysical
$b_\pm$ ($b_-<0$ in part). The numerical results below show that this does not
happen and confirm the conjectured
form of the flame profile.\footnote{To understand better
the difference between the situation here and that in the previous section one
may start with some physical ($=\,$satisfying b.c., $f(-\infty)=1$)
$f$ on the left and check if it can satisfy $f(+\infty)=0$ as well.
As was mentioned in
footnote (15) there is a unique solution (modulo translations) of the
linearized equation with step-function $\Phi_0$, which goes to zero
as $\chi\to+\infty$. To eliminate a possible objection that the
non-zero $c_1$ makes predictions based on solutions of linearized equation
questionable it is instructive to find a general
solution going asymptotically to $c_1$: this reads
\begin{equation}\label{fc1}
f=c_1+2\Lambda\left(\Bigl(1+\frac{2\Lambda}{f_0-c_1}\Bigr)e^{d(1+\frac{c_1}\Lambda)\chi}-
1\right)^{-1}
\end{equation}
%$\forall c_1\in[0;f_0]$%for which such solutions exist
 (up to translations in $\chi$). As $c_1$ increases from 0 to $f_0$
$\:\left.\frac{df}{d\chi}\right|_{f=f_0}$ increases monotonically from
$-f_0d\left(\frac{f_0}{2\Lambda}+1\right)$ to 0. If for a given $d$ the
(unique up to translations) $f$ going to 1 as $\chi\to-\infty$ has
$\left.\frac{df}{d\chi}\right|_{f=f_0}\in[-f_0d(\frac{f_0}{2\Lambda}+1);0]$
 it will go asymptotically to the
corresponding $c_1$ as $\chi\to +\infty$, namely it will be exactly
\p{fc1} where $f<f_0$; if its derivative is more negative, it will be
described by \p{fc1} with negative $c_1$ until it intersectss $f=0$ at
some finite $\chi$ (and this is not a solution we are interested in).
Easily established monotonousness properties prove this
way the claim that there is a unique $d$ for which the physical
solution (according to its behavior near $f=1$) goes asymptotically
to zero as $\chi\to-\infty$.

For the $\Phi(f)$ \p{Phikolm}, on the other hand, there are \emph{no} solutions
going asymptotically to any constant but 0 as $\chi\to+\infty$, but those going
to 0 represent a two-parameter set of solutions, and depending on the $b_\pm$ in
their asymptote the $\left.\frac{df}{d\chi}\right|_{f=f_0}$ may take different
values, thus making it possible to match the
$\left.\frac{df}{d\chi}\right|_{f=f_0}$ of the solution physical near $f=1$ for
a range of $d$\rq s. In this case $f_0$ denotes some intermediate value
of $f$, between 0 and 1.}\\
\indent 3) $d=2$: \p{krightasym} must be rewritten as
$f(\chi)=(b+b_0\chi)e^{-\frac d 2 \chi}$, and there is again a 2D domain of
physical $(b,b_0)$, yielding $f>0$, $f\to 0$ as $\chi\to+\infty$. In this case a
meaningful burning profile exists as well.

Analysis in this section can be straightforwardly generalized as follows:\\
%\indent i) For any $\Phi_0(f)>0$ on $(0;1)$ and going to 0 both at $f\to 0$ and
%$f\to 1$ with non-zero derivatives there is a region of $d$ values, $d\in[2\sqrt
%\omega;+\infty)$, for which eigenfunctions exist. Here
%$\omega=\frac{d\Phi_0}{df}\bigr|_{f=0}.$\\
%
\indent i) If $\Phi_0(f)$ goes to some (positive) constant as $f\to 0$ there
are no solutions going to 0 as $x\to+\infty$. From the physical viewpoint the
system reacting with finite rate in the initial state is unstable and
self-similar solutions cannot exist. Interestingly, some problems with
$\Phi_0(0+)=\nu<0$ do have eigenvalues. Corresponding eigenfunctions are identically zero
on the right of some $\chi_2$, $f(\chi)\sim\bigl(\nu(\chi-\chi_2)/2\bigr)^2$ as
$\chi\to\chi_2-$.\\
\indent ii) For $\Phi_0(f)=\mu f +o(f)$ at $f\to 0$ non-negative solutions going to 0 as
$x\to +\infty$ exist iff $d\ge 2\sqrt{\mu}$, $\mu>0$ for positiveness
of $\Phi_0(f)$. General solution getting to a vicinity of  $f=0$ decays
exponentially. %\\ \indent $\hphantom{\\iii}$
Analysis near $f=1\Leftrightarrow\bar{f}=1-f\ll 1$ then
suggests the following for
$\Phi_0=\bar{\nu}+\bar{\mu}\bar{f}+o(\bar{f})$ (below $d\ge 2\sqrt{\mu}$
is assumed; $f(-\infty)=1$ required): \\
\indent\hspace{3mm}$\bullet$\ $\bar{\nu}>0,\bar{\mu}> 0$ -- for all $d$
(the above $d\ge 2\sqrt{\mu}$ assumed) a unique
profile exists and $\exists x_1: \forall x<x_1\; f(x)\equiv 1$.
% is again identically 1 on a half-line.\\
%%
\indent\hspace{3mm}$\bullet$\ $\bar{\nu}>0,\bar{\mu}\le 0$ -- a unique
profile whenever $\bar{d}\equiv d(1+\frac 1\Lambda)\ge 2\sqrt{-\bar{\mu}}$
(i.e. for $|\bar{\mu}|>\mu(1+1/\Lambda)^2$ the spectrum is additionally shrinked);
again identically 1 on a half-line.
%;$\exists x_1: \forall x<x_1\;
%f(x)\equiv 1$.
\\
\indent\hspace{3mm}$\bullet$\ $\bar{\nu}=0,\bar{\mu}> 0$ -- a unique
solution exponentially approaching 1 as $x\to-\infty\;\forall d$.\\
\indent\hspace{3mm}$\bullet$\ $\bar{\nu}=\bar{\mu}=0$. If
$\Phi_0(f)=\bar{q}\bar{f}^r+o(\bar{f}^r)$, $r>1$,\ $\forall d$ a unique solution
exists. %, slowly approaching $f=1$.
The solution of \p{master1}
may be written as
$p=\frac{\bar{q}}\bd \fb^r\bigl(1+\frac \fb{\Lambda+1}-\frac{r\bar{q}}{\bd^2}\fb^{r-1}+
O(\fb^2+\fb^r)\bigr)$,
leading to $\fb\sim\bigl((1-r)\bar{q}\chi/\bd\bigr)^{\frac 1{1-r}}$.
%%
%%\indent\hspace{3mm}$\bullet$\
If $\Phi_0(f)\equiv 0$ in some neighborhood of $f=1$ bounded solutions exist
but $f\to 1$ at $x\to -\infty$ \emph{from above}. Not catastrophic
for FCT, but unpleasant.\\
\indent\hspace{3mm}$\bullet$\ $\Phi_0\sim\bar{q}\bar{f}^r$, but $r\in(0;1)$.
$\fb$ vanishes identically on the left of some $\chi_1$, in its right neighborhood
$\fb\sim\bigl(\sqrt{\bar{q}/2(1+r)}(1-r)(\chi-\chi_1)\bigr)^{\frac 2{1-r}}$.\\
\indent iii) For $\Phi_0(f)=qf^r$ near $f=0$ ($q>0$) there are no solutions of \p{master1}
with $p(+0)=0$ when $r<1$ -- a situation analogous to i). When $r>1$ on the other hand,
there are multiple solutions going to 0 (the ODE\rq s peculiarity), hence one expects the same behavior
as in ii), depending on the $\Phi_0$ shape near $f=1$. When $\Phi_0(f)\equiv 0$ in some
neighborhood of $f=0$ one would expect a discrete spectrum of $d$, with corresponding
eigenfunctions behaving near $f=1$ according to the $\Phi_0(f)$ near 1 as in ii).

Summing this up, the KPP-like behavior is quite universal for $\Phi_0(f)$
going to 0 linearly or faster at $f\to 0$; on the other hand,
$\Phi_0(f)\equiv 0$ near $f=0$ leads to the spectrum similar to that in the
previous section; positive $\Phi_0(f)$ decreasing slower than linearly
(or not going to zero) as $f\to 0$ lead to absence of steady-state solutions.
Profile $f(x)$ has an infinite tail at $f=1$ iff $\Phi_0|_{f\to 1}$
decreases linearly or faster.

\begin{figure}[htbp] \begin{centering}%\vspace*{-4cm}\hspace*{-20mm}
\includegraphics[width=0.5\textwidth]{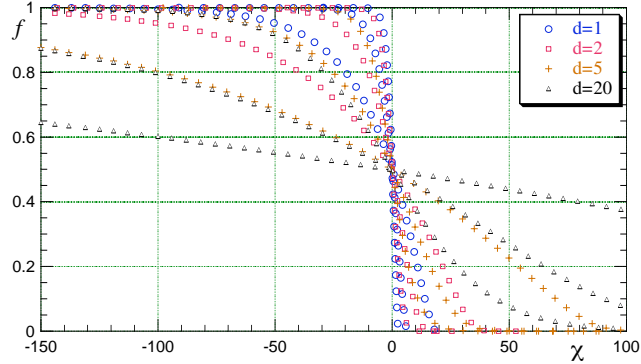}
\caption{Flame profiles at different $\Lambda$ and $d$.
For each $d$ 3 curves are depicted, for $1/\Lambda=0.05,\,5,$ and 20,
larger $1/\Lambda$ corresponding to the curves
intersecting $\chi=0$ axis (where $f$ was set to 0.5) at larger
angles, and having larger widths.}
\end{centering}\end{figure}

Flame profiles $f(\chi)$ found numerically for $\Phi_0(f)=f(1-f)$
are shown in Fig.~2 for four values of $d$. These seem to satisfy the boundary
conditions for $d\ge 2$, whereas the profiles for $d=1$ (each integrated
from $p(1)\equiv -df/d\chi|_{f=1}=0$) intersect $f=0$ at
finite $\chi$ with non-zero $df/d\chi$. Corresponding integral curves
$p(f)$ at the same values of $d$ and $\Lambda$ are represented in Fig.~3.
Note that for $d=1$ \ $p|_{f=0}\ne 0$, whereas for $d\ge 2$ it was
checked that by refining the grid $p(0)$ became correspondingly closer to 0,
down to $p(0)\approx 10^{-10}$ at the bulk grid spacing
of $10^{-7}$ (at finer grid rounding errors in real(8) dominated).
This numerically confirms our qualitative conclusions.

Note that the flame width grows fast with $d$. As in the original
model \cite{KPP} (only asymptotes near $f=0$ are essential for the
argument to hold, and these have the same exponential form regardless of $\Lambda$)
one may conclude that only propagation with the smallest velocity is
stable. Of course, if the $f(x)$ at some time corresponds to some
eigenfunction above, such a profile will propagate with
corresponding $d$. But if one considers a process of setting up the
steady-state propagation, with initial $f(x)$ corresponding to pure
fuel on a half-line (or $f$ decaying with $x$ faster than the fastest
exponent $\lambda_-(d=2)=1$ in \p{krightasym}) the conclusion
will be that the resulting self-similar front will be the
eigenfunction with the smallest velocity. More generally, by
considering the evolution of some superposition $\Psi$ of the
steady-state profiles found above one concludes that amplitudes of
all of them but one will (asymptotically) decrease in time in favor of the one with
the smallest $d$ in the spectrum of $\Psi$ (they interact due to
nonlinearity of the system), decay rate being proportional to
difference of corresponding $\lambda$\rq s. As a generic
perturbation is a superposition of all the eigenfunctions, however
small it is it will eventually reshuffle the profile into that for
smallest $d$.

\begin{figure}[htbp] \begin{centering}%\vspace*{-55mm}\hspace*{-51mm}
\includegraphics[width=0.5\textwidth]{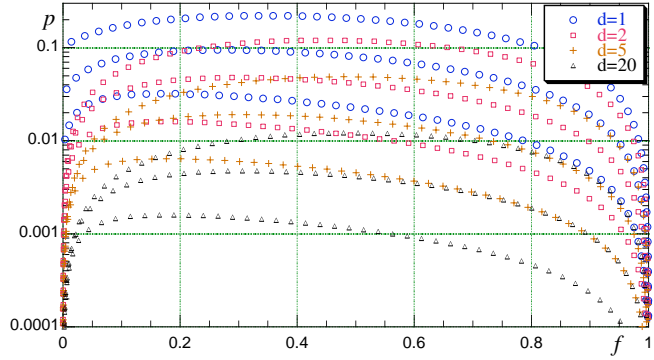}
\caption{$p(f)$, integrated from $f=1,\, p=0$ at $\Lambda$ and $d$ as in Fig. 2; larger
$1/\Lambda$ correspond to smaller $p$.
Note that at $d=1$ $p(f=0)$ is non-zero, in contrast with the $d\ge 2$ curves.}
\end{centering}\end{figure}

 It does not seem feasible to use KPP-like profile in FCT
because of the continuous spectrum of the velocities, thus long times of
relaxing to steady state and large widths with two exponential tails.
%%%%%%%%%%%%%%%%%%%%%%%%%%%%%%%%%%%%%%%%%%%%%%%%%%%%%%%%%%%%%%%%%%%%%%%%%%%%
\section{Step-function: velocities and widths}
\subsection{Analytic solution. Model I}
Let us fix the solution (thus eliminating the translational invariance) such
that $f(0)=f_0$. By going to dimensionless
\begin{equation}\label{rescaledvar}
\xi=d\chi=\frac{D_f}K x,\;\phi=f/\Lambda
\end{equation}
and integrating once over $\xi$ one gets the first order equation for
$\phi(\xi )$:
%\footnote{we consider \p{rho1} model now; the \p{rho2} one is
%studied right afterwards.}
\begin{equation}\label{rescaledmaster}
\frac{d\phi}{d\xi}+\phi\left(1+\frac{\phi}
2\right)+\frac{\Phi_0\xi}{d^2\Lambda}=A.
\end{equation}
Somewhat frivolous notation allows using this for two regions: where
$\phi\le f_0/\Lambda$ (using $\Phi_0=1$) and where
$f_0/\Lambda<\phi<1/\Lambda$ ($\Phi_0=0$). $A$ is an integration constant,
a priori different for these two regions. The boundary condition
$\phi=d\phi/d\xi=0$ at $\xi=+\infty$ fixes $A=0$ for the right
half-line ($\xi\ge 0$); one can further see that for model I left $A$
is zero as well, by enforcing continuous
differentiability of $\phi(\xi)$ at $\xi=0$.

One can thus evaluate the flame width as (in the original units)
\begin{equation}\label{w1}
W_1=\frac {f_0}{\left|df/dx\right|_{f=f_0}}=\frac
K{D_f}\left(1+\frac{f_0}{2\Lambda}\right)^{-1}= \sqrt{\frac K
R}\left(d\left(1+\frac{f_0}{2\Lambda}\right)\right)^{-1}\equiv \sqrt{\frac K
R}w_1,
\end{equation}
as well as the $\xi_1$ ($\xi_1\equiv \chi_1/d$ is the rightmost point
where $\phi=1/\Lambda$: $\phi^{-1}(\lbrace
1/\Lambda\rbrace)=(-\infty;\xi_1]\,$)

\begin{equation}\label{xi1}
\xi_1=-d^2\left(1+\frac{1}{2\Lambda}\right).
\end{equation}
Note that the solution at positive $\xi$ is (in the original variables)
\begin{equation}\label{sol2}
f=2\Lambda\left(\Bigl(1+\frac{2\Lambda}{f_0}\Bigr)e^{\frac{D_f}K x}-1\right)^{-
1}\approx\Bigr(\frac 1{2\Lambda}+\frac 1{f_0}\Bigr)e^{-\frac{D_f}K x},
\end{equation}
the last estimate is for $|x|\gg\bigl|\frac K{D_f}\bigr|\ln
%\bigl(\frac{2f_0}{\Lambda}+\frac{f_0^2}{\Lambda^2}\bigr)$
\bigl(1+{2\Lambda}/{f_0}\bigr)$; thus for small $d$ (achieved, say,
at small $\Lambda$ -- see below) the region where solution decays slower
than asymptotic exponential is wide (may be wider than $w_1$ at
physically reasonable parameter values), and the definition of width
is ambiguous (more so taking into account complicated
$f(x)$ dependence on the left half-line, cf.~\p{genphi}). Therefore
we will employ another definition for comparison
\begin{equation}\label{w2}
w_2=\frac{\chi(f=0.1)-\chi(f=1)}{0.9}
%{x|_{f=0.1}^{1}}\Bigm/{0.9}
\end{equation}
in numerical estimates. Analytically one finds (for $f_0\ge f_-\equiv 0.1 $)
\begin{equation}\label{w2exp}
w_2=\frac{1}{d}\left(\ln \frac{1+2\Lambda/f_-}{1+2\Lambda/f_0}
+d^2\Bigl(1+\frac{1}{2\Lambda}\Bigr)\right),
\end{equation}
the first term derived from the form \p{sol2} and the second one from the above
$\xi_1$.\\

In the region $(\xi_1,0)$ one can express general solution of
\p{rescaledmaster} in terms of Airy (or Bessel) functions by substitution
\begin{equation}\label{yphi}
\phi=-1+\frac 2 y \frac{dy}{d\xi};
\end{equation}
one gets\footnote{this substitution does not introduce new
integration constants, as rescaling of $y$ (the resulting equation
for $y(\xi)$ is linear) leaves $\phi$ unchanged. This was used
in \p{genphi}, where the coefficient at $\mathrm{I}$ function was
set to unity. This is possible for all the solutions $y$
except $y=\sigma^{1/3}\cdotp B\mathrm{K}_{1/3}(\sigma)$; the
latter never yields eigenfunctions in our problem%
%(but generally one should think of $B\in\mathbb{RP}_1$.)
.}
\begin{equation}\label{genphi}
\phi=-1-\left(\frac{6\sigma}{d^2\Lambda}\right)^{1/3}\left(\frac{1}{3\sigma}+
\frac{d}{d\sigma}\ln\left(\mathrm{I}_{1/3}+B\mathrm{K}_{1/3}\right)(\sigma)
\right).
\end{equation}
Here
\begin{equation}\label{sig}
\sigma= \left(\frac 2\Lambda \right)^{1/2}\left(\xi-
\frac{d^2\Lambda}2\right)^{3/2}
\end{equation}
and $\mathrm{I},\k$ are modified Bessel functions. By imposing boundary conditions
\begin{equation}\label{bdy}
\phi=\left\lbrace\begin{array}{ll}
f_0/\Lambda & \mbox{at } \sigma_0=d^2\Lambda /6\\
1/\Lambda & \mbox{at } \sigma_1=\left(1+\frac 1\Lambda\right)^3\sigma_0
%\equiv k\sigma_0
\end{array}\right. ,
\end{equation}
one can get the following equation, determining eigenvalues of $d$:
\begin{equation}\label{eigd}
\left(\mathrm{I}_0\Bigl(1+\frac{f_0}\Lambda+\frac
1{3\sigma_0}\Bigr)+\mathrm{I}_0^\prime\right)\left(\k_1\Bigl(1+\frac
1{3\sigma_1}\Bigr)+\k_1^\prime\right)-
%\left(\k_0\Bigl(1+\frac{f_0}\Lambda+\frac
%1{3\sigma_0}\Bigr)+\k_0^\prime\right)\left(\mathrm{I}_1\Bigl(1+\frac
%1{3\sigma_1}\Bigr)+\mathrm{I}_1^\prime\right)=0.
\mathrm{I}\leftrightarrow\k=0.
\end{equation}
The notation is $\mathrm{I}_0=\mathrm{I}_{1/3}(\sigma_0)$, $\k_1=\k_{1/3}(\sigma_1)$, etc.; the
second term differs from the first one by switching functions
$\mathrm{I}\leftrightarrow\k$, i.e. it reads
$\Bigl(\k_0\bigl(1+\ldots\bigr)
+\k^\prime_0\Bigr)\Bigl(\mathrm{I}_1\ldots \Bigr)$. With the above definitions this
equation has a unique solution for $d$ for every $f_0,\:\Lambda$.

\subsection{Model II}
\begin{figure}[htbp] \begin{centering}%\vspace*{-4cm}\hspace*{-20mm}
\includegraphics[width=0.6\textwidth]{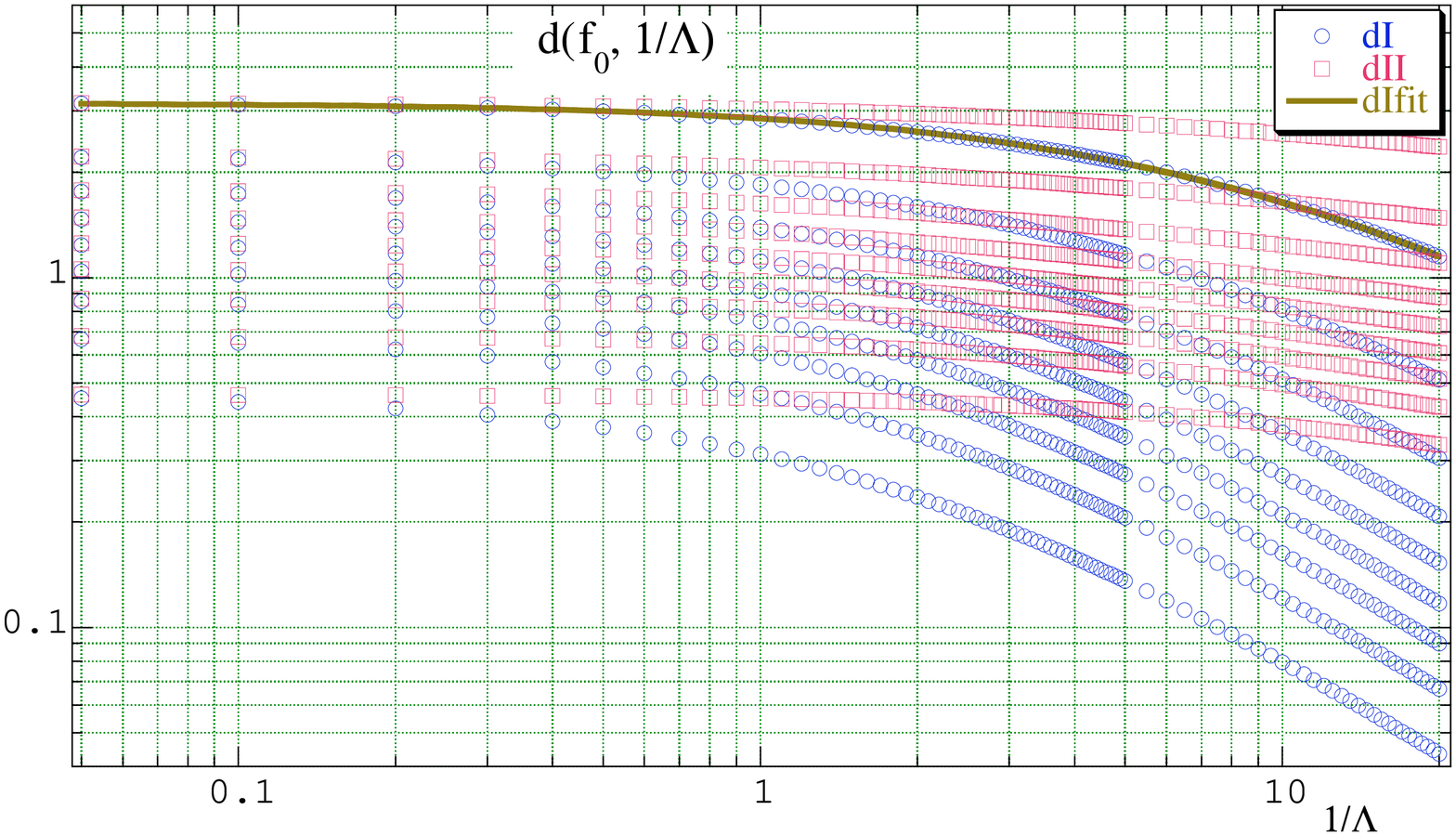}
\caption{Flame velocities. The nine sequences
$d(\Lambda)$ for each model (blue circles and red squares for model I and II respectively)
correspond (top to bottom) to $f_0=0.1,$\ldots 0.9 with step 0.1. The worst
$(f_0$=0.1) fit (\ref{mf0}) for model I is shown.}
\end{centering}\end{figure}

In a similar way one may analyze the boundary problem with $\rho$
given by \p{rho2}. ($\tilde{D}_f$ below denotes the eigenvalue of
(\ref{artvisfront0}) with (\ref{bndl}--\ref{bndm}),
$\tilde{d}$ -- corresponding dimensionless speed, e.v. of \p{artvisfront0}.)
One gets exponential decay of $f$ in $(0;f_0)$, $f=f_0e^{-\frac{\tilde{D}_f}K x}$, the widths being
\[ \tilde{W}_1=\textstyle{K/{\tilde{D}_f}},\quad \tilde{W}_2=
\frac{K}{\tilde{D}_f}\left(\ln
\frac{f_0}{f_-}+\tilde{d}^2\left(1+\frac{1-f_0}{2\Lambda}\right)\right).
\]
The equation for \[ \tilde{\phi}(\xi)=\frac{f(\xi)-f_0}\tl\equiv
\frac{f(\xi)-f_0}{\Lambda(1-f_0)} \]
in the region $(\tilde{\xi}_1;0)$,
\[ \frac{d\tilde{\phi}}{d\xi}+\tilde{\phi}\Bigl(1+\frac{\tilde{\phi}}2\Bigr)+
\frac{\Phi_0\xi}{d^2\tl}+\frac{f_0}{\tl}=0, \] again has a general
solution in terms of Airy functions\footnote{this form is more
convenient here, as the argument $s$ of Airy functions $\a$ and $\b$
(both bounded at 0) may change sign on $(\tilde{\xi}_1;0)$. In
solving the equation \p{tilded} the Airy functions normalization from
\cite{numfor} was used.}:
\begin{equation}\label{gensol2}
\tilde{\phi}=-1-(\tilde{d}^2\tl/4)^{-1/3}\frac{d}{ds}
\ln\left(\a+\tilde{B}\,\b\right)(s),
\end{equation}
which after imposing the boundary conditions as above yields the equation for
$\tilde{d}:$
\begin{equation}\label{tilded}
\left(\b_1\sqrt{s_1}+\b^\prime_1\right)\left(\a_0\sqrt{\frac{s_0}{1-2f_0/\tl}}+
\a^\prime_0\right)-\a\leftrightarrow\b=0.
\end{equation}
In the above \[
s=-(2\tilde{d}^2\tl)^{-1/3}\biggl(\xi-\frac{\tilde{d}^2\tl}{2}(1-
2f_0/\tl)\biggr), \] and subscripts at functions again represent
arguments (subscript \lq\lq 0\rq\rq\ means argument
$s_0=(\tilde{d}^2\tl/4)^{2/3}(1-2f_0/\tl)$, \lq\lq 1\rq\rq\ similarly implies
$s_1=(\tilde{d}^2\tl/4)^{2/3}\bigl(1+\frac{1-f_0}{\tl}\bigr)^2$).

\subsection{Results for velocities and widths}

\begin{figure}[htbp] \begin{centering}%\vspace*{-4cm}\hspace*{-20mm}
\includegraphics[width=0.5\textwidth]{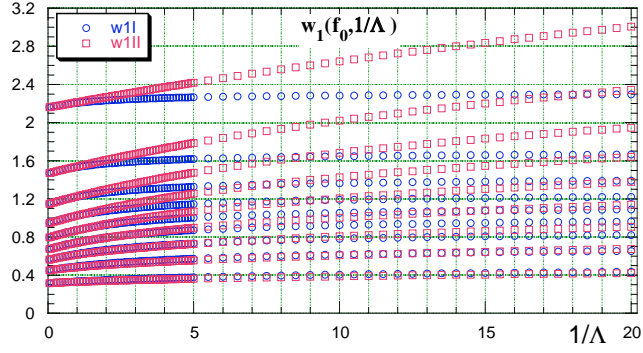}
\caption{Flame width $w_1$ at the same $\Lambda$ and $f_0$.
Larger widths correspond to larger $f_0$.}
\end{centering}\end{figure}

Numeric solutions of \p{tilded} and of \p{eigd} are presented in Fig.~4.%
\footnote{Possibility of negative arguments for Airy functions in model II
leads to infinite sequence of solutions of \p{tilded} for the $d$. However,
only the $1^\mathrm{st}$ (smallest) $d$ corresponds to a legitimate
eigenfunction: the \lq\lq eigenfunction\rq\rq\ $f_n(\chi)$ corresponding to
the $n^\mathrm{th}$ root for $d$ has $n-1$ first-order poles (but it
perfectly goes to 0 as $\chi\to\infty$ and $f(\chi)=1\,\forall\chi<\chi_{1n}$.)}
Figs.~5 and 6 show the
$w_1$ and $w_2$ for these two models.
Flame profiles in Fig.~1 were obtained by direct numerical integration of
\p{master12} with the $d$ from Fig.~4. Relative difference between
$d$\rq s found numerically and analytically is less than
$5\times 10^{-3}$\%. It was that large for $f_0\ge 0.7$;
for $f_0\in[0.3; 0.6]$ the difference was of order $10^{-8}$, the accuracy
required of the $d$ in numeric procedure (that for solving \p{eigd} and \p{tilded}
was set to
$4\times 10^{-16}$ and apparently these errors played no role), and monotonically
increased to $20\times 10^{-8}$ with $f_0$ further decreasing to $0.1$.
Quite similar numerical integration errors were observed for model II,
reaching $8.4\times 10^{-3}$\% (again, with bulk integration step in $f$
being $10^{-5}$). These largest errors dropped to $10^{-3}$\% when the bulk
integration step was 8 times decreased. Errors in widths followed similar
trends (and \lq\lq numerical\rq\rq\ widths were consistently larger than
\lq\lq analytical\rq\rq\ ones, whereas velocities were smaller), though
there was additional contribution from crude estimation (up
to the whole grid) from integral curves $x(f)$ obtained from $p(f)$ via
further trapezium integration; the discrepancy in $w_{1,2}$ in both models
was less than $8\times 10^{-3}$\%.

\begin{figure}[htbp] \begin{centering}%\vspace*{-4cm}\hspace*{-20mm}
\includegraphics[width=0.5\textwidth]{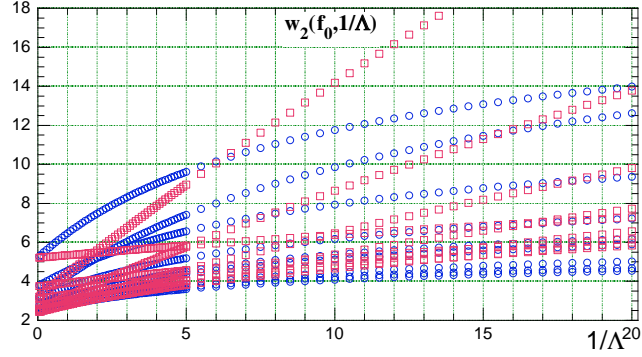}
\caption{Flame width $w_2(f_0,\Lambda)=-\chi|_{f=0.1}^1 \Bigm/0.9$. For model I
the order of $f_0$ (from larger to smaller $w_2$ at $1/\Lambda=20$) is 0.9,
0.1, 0.8,
0.7, 0.2, 0.6, 0.5, 0.3, $0.4\,$. That for model II is 0.1, 0.2, 0.3, 0.4,
0.9, 0.5, 0.8, 0.6, $0.7\,$.}
\end{centering}\end{figure}

For the problem of immediate interest (deflagration in SN Ia) matter
expansion is not large, thus it is worth trying to treat $1/\Lambda$ as a small
parameter. Resulting asymptotic expansion looks most natural in $h=1/d^2$:
{\setlength\arraycolsep{2pt}
\begin{eqnarray}\label{Lambdanearinf}\nonumber
(f_0-h)e^{1/h}+h&-&\frac 1{2\Lambda}\left[4h+5h^2+e^{1/h}
\left((f_0-h)^2-5h^2-2(f_0-h)\Bigl(1+\frac 1h\Bigr)\right)\right.\\
  &-&\left.\vphantom{e^{2/h}\frac 1 h}
(f_0-h)^2e^{2/h}\right]+O(\Lambda^{-2})=0,
\end{eqnarray}}%
from which one finds a first-order correction to the solution $h_0\equiv d^{-2}$
of \p{Lambdainf}, 
\begin{equation}\label{dasym}
h=h_0\left(1+\frac{h_1}{2\Lambda}+O(\Lambda^{-2})\right),\; 
h_1=5h_0-e^{-1/h_0}\frac{2+h_0-h_0f_0}{f_0-e^{-1/h_0}}.
\end{equation}
When $f_0$ is small as well this may be estimated with the aid of
expansion from the end of Sec.~3.1 for $h_0$:
\begin{equation}\label{dasymf0}
h_1=5f_0-e^{-1/f_0}\left(\frac 2{f_0}+1-6f_0\right)
-e^{-2/f_0}\left(\frac 4{f_0^2}+\frac 2{f_0\vphantom{f_0^2}}-6-6f_0
\right)+O(e^{-3/f_0}/f_0^3),
\end{equation}
this is usable up to $f_0\approx 0.3$.
Error of using \p{dasym} does not exceed 1\% for $1/\Lambda\le 1$ (but $d=h^{-1/2}$ must be used.
Expanding $d$ up to $O(1/\Lambda)$ leads to far worse agreement.) At $1/\Lambda=3$ the error
grows to 6.7\% for $f_0=0.3$ (1.8\% for $f_0=0.2$); at $\Lambda>0.69$ (SN Ia problem) it is
within 2.1\% for $f_0=0.3$, 0.16\% for $f_0=0.2$.
At these $f_0$ both expressions for $h_1$ give the same agreement.

A very similar $1/\Lambda$ correction to \p{Lambdainf} may be written for model
II by observing that factors in front of $\tl^0$ and $\tl^{-1}$ under $1/\tl$
decomposition of \p{tilded} (rewritten in terms of $\mathrm{I}$ and $\k$ functions)
coincide with
the factors at the corresponding powers of $\Lambda$ in the decomposition of
\p{eigd}; therefore (\ref{Lambdanearinf}--\ref{dasymf0}) with $\Lambda$ switched to $\tl$
provide corresponding estimates of $\tilde{d}(f_0,\tl)$.
%%%%%%%%%%%%%%%%%%%%%%%%%%%%%%%%%%%%%%%%%%%%%%%%%%%%%%%%%%%%%%%%%%%%%%%%%%%

\subsection{Finite width flames}
As we saw in the previous section an important drawback of KPP-like burning
rates is related to exponential tails in the corresponding
eigenfunctions $f(\chi)$. Behavior of the FCT models based on $\Phi_0(f)$
vanishing identically near $f=0$ in non-stationary simulations
is better as they admit only one steady-state solution, yet for these
$f(x)$
still has an infinite tail thus creating unphysically wide preheating
zone in simulations hindering performance in the presence of large gradients.
In this section we modify the \emph{diffusion} term in \p{artvis0} to
make the total flame width finite.

Several possibilities have been investigated,
\begin{equation}\label{artvisK}
\frac{\partial(\rho f)}{\partial t}+\frac{\partial(\rho fu)}{\partial x}= \rho
\frac{\partial}{\partial x}\left(\tilde{K}K_0(f)\frac{\partial f}
{\partial x}\right)+\rho R\Phi_0(f)
\end{equation}
with $K_0(f)=f^r$, $r\in(0;1)$ seems the simplest one providing
promising results.\footnote{%
It might seem more physical to write the diffusion
term as $\nabla(\rho K\nabla f)$. This case was studied as well, normalized
to unit total width flame profiles are more sensitive to $\Lambda$ than the ones
corresponding to \p{artvisK}; such a model is also less tractable
analytically. The model with diffusion term of this form and with $K(f)=const$
was also studied as a possible alternative to the original one \p{artvis0}.
Asymptotic behavior of $d$ and the widths at small and large $\Lambda$ is the same as
presented in the previous section, hence the decision to stick to the original
(apparently consistently performing) model.

As with $\Phi_0(f)$ we are free to choose from a variety of diffusion-like terms;
an argument favoring the \lq\lq most physical\rq\rq\ form discussed in
this footnote might be its more natural response to nonstationary hydrodynamic
factors (following the logic of Introduction). However, physical transport
coefficients change drastically after passing through the flame, making it
questionable if putting the $\rho$ under $\nabla$ is really an improvement, in view
of the artificial $K_0(f)$ dependence chosen vs effective $K(f)$ for turbulent burning.}
With $\rho_0/\rho=1+f/\Lambda$ this problem admits a unique steady-state solution
with $D_f=d\sqrt{\tilde{K}R}$, $d$ being the eigenvalue of
\begin{equation}\label{master1K}
\frac{d}{d f}\bigl( K_0(f)p\bigr)-d\Bigl(1+\frac f\Lambda\Bigr)+\frac{\Phi_0(f)}p=0
\end{equation}
with $p(0)=p(1)=0$ (as before $p=-df/d\chi\equiv -\sqrt{\tilde{K}/R}\,df/dx$).
The above $K_0(f)$ leads to $f(\chi)$ $C^1$-smoothly monotonically interpolating
between $f=1\: \forall \chi<\chi_1=-d(1+1/2\Lambda)$ (fixing $f(0)=f_0$;
$f=1-(\chi-\chi_1)^2/2+d(\chi-\chi_1)^3/2+\ldots$ at $\chi\to\chi_1+$) and
$f=0\:\forall \chi>\chi_2$ ($f=(rd(\chi_2-\chi))^{1/r}(1+O(\chi_2-\chi))$ at
$\chi\to\chi_2-$).

$\chi_2$ and the total width $w_3=\chi_2-\chi_1$ may be expressed in elementary
functions of $f_0$, $\Lambda$ and $d$ for rational $r$ (or as incomplete
$\Gamma-$function in general). Values $r=1/2$ and $r=3/4$ seem most adequate.
Corresponding widths are
{\setlength\arraycolsep{2pt}
\begin{eqnarray}\label{w312}
w_{3,\frac 12}&=&\frac 1d\int_0^{f_0}f^{r-1}\left(1+\frac f{2\Lambda}\right)^{-1}
\,df\biggr|_{r=1/2}-\chi_1=\frac 2d\sqrt{2\Lambda}\arctan\sqrt{\frac{f_0}{2\Lambda}}+d\left(1+\frac 1{2\Lambda}\right)\\ \label{w334}
w_{3,\frac 34}&=&2^{5/4}\Lambda^{3/4}d^{-1}\left[\textrm{Arctan}\,\frac{(2f_0/\Lambda)^{1/4}}
{1-(f_0/2\Lambda)^{1/2}} +\ln\frac{(f_0/2\Lambda)^{1/2}-(2f_0/\Lambda)^{1/4}+1}
{(f_0/2\Lambda+1)^{1/2}}\right]+d\left(1+\frac 1{2\Lambda}\right)
\end{eqnarray}}%
In the last expression the branch of Arctan used is Arctan$\,\in[0;\pi)$.

For any $r$ in the limit $\Lambda\to\infty$
{\setlength\arraycolsep{2pt} \begin{eqnarray*}
w_{3,r}&=&\frac{f_0^r}{dr}\left(1-\frac{f_0}{2\Lambda}\,\frac r{r+1}+\Bigl(\frac{f_0}{2\Lambda}\Bigr)^2\,\frac
r{r+2}+O\Bigl(\frac{f_0}{2\Lambda}\Bigr)^3\right)+d\Bigl(1+\frac 1{2\Lambda}\Bigr)\\
 &=&\Bigl(\frac{f_0^r}{rd_0}+
d_0\Bigr)+\frac 1{2\Lambda}\left(d_0\Bigl(1-\frac{h_1}2\Bigr)-\frac{f_0^r}{rd_0}\Bigl(\frac{f_0r}{r+1}-
\frac{h_1}2\Bigr)\right)+O(\Lambda^{-2})
\end{eqnarray*}}%
(as in the $r=0$ case $d(f_0,r|\Lambda)\equiv h^{-1/2}=d_0(1+\frac{h_1}{2\Lambda}+\ldots)^{-1/2}$;
$d_0=h_0^{-1/2}$, $h_1$, etc. are now some functions of $f_0$ and $r$);
as $\Lambda\to 0$ $w_3$ diverges as
\[
w_{3,r}=\frac{(2\Lambda)^r}{d}\,\frac{\pi}{\sin\pi r}-\frac{f_0^{r-1}}{1-r}\,
\frac{2\Lambda}d+O(\Lambda^r)-\chi_1=\frac{2^r\pi}{G_0\sin\pi r}\Lambda^{r-1}(1-G_1\Lambda)
+\frac{G_0}2-\frac 2{G_0}\,\frac{f_0^{r-1}}{1-r}+O(\Lambda).
\]
We used $d=G_0\Lambda(1+G_1\Lambda+o(\Lambda))$ as in the step-function model with a standard diffusion term:
$d(\Lambda)$ dependence is qualitatively very similar. The fits of $d(\Lambda)$ for $r=3/4$
and various $f_0$ are presented in the next section.

\begin{figure}[htbp] \begin{centering}%\vspace*{-4cm}\hspace*{-20mm}
\includegraphics[width=0.9\textwidth]{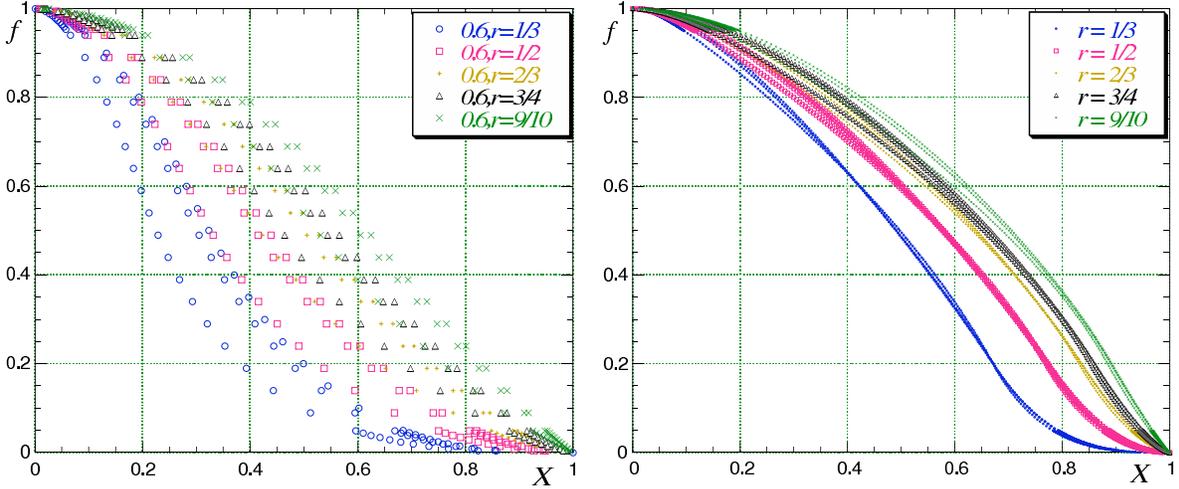}
\caption{Normalized to unit width flame profiles with $f_0=0.6$ (left)
and $f_0=0.2$ (right). The three curves for each $r$ correspond to $1/\Lambda\in\lbrace
0.16;0.6;3\rbrace$, lower near $f=0$ profiles for larger $1/\Lambda$.}
\end{centering}\end{figure}

For a range of $\{f_0,r\}$ flame profiles deliver what they were expected to originally.
Their width $w_3$ may consistently serve as a measure of $f$ gradients, and upon coupling \p{artvisK} to
the hydrodynamic equations one would really get reasonably uniform heat release within the flame
width. On the contrary, for the models with traditional diffusion term $w_2$ was somewhat
arbitrary quantity, most of it (for larger $f_0$ and KPP to a greater degree) might correspond to
\lq\lq tails\rq\rq\ of a profile, and the heat release in FCT would remain localized, contrary to
the intention to more-or-less uniformly distribute it over a few cells near the modelled flame front.
Some representative flame profiles are shown in Fig.~7; they are normalized to unit width (that is,
reexpressed in terms of a new $X=(\chi-\chi_1)/(\chi_2-\chi_1)$; resulting
supp$(df/dX)=[0;1]$); they illustrate our \lq\lq top 3 picks\rq\rq\ to be used in SN Ia
simulations:\\
\indent\hspace{3mm}$\bullet$\ $r=1/2$. This has advantage that $f(x)$ behavior near
$f=0$ is the same as near $f=1$, thus we are unlikely to introduce new problems
(compared to the original $r=0$ scheme). $f_0=0.2$ is then convenient as the $f(x)$ shape
is least sensitive to $\Lambda$ in its range of immediate interest, $1/\Lambda\in[0;2]$.\\
\indent\hspace{3mm}$\bullet$\ $r=3/4$, $f_0=0.6$. For the $\Lambda$\rq s of interest corresponding
flame profiles seem most symmetric overall w.r.t. $f\mapsto 1-f$; this is perhaps a
better realization of the above idea: as the whole $w_3$ width will be modelled by 3--4
integration cells this approximate global symmetry seems more adequate to consider than
the symmetry of minute regions near $f=0$ and $f=1$.\\
\indent\hspace{3mm}$\bullet$\ $r=3/4$, $f_0=0.2$. Profile gradients are still uniform enough
on $[0.1;0.9]$, they drop to zero in a regular manner within reasonable $\Delta\chi\leq 0.25w_3$.
At $f_0=0.2$ the profiles seem least sensitive to $\Lambda$, ensuring consistent performance in all
regions of the star. 
%My impression is that this ${K_0,\Phi_0}$ combination is the most promising one.\\
\section{Implementation suggestions}

Analytically found asymptotes for $d(\Lambda)$ (Sec.~4.3) at small $1/\Lambda$
do not provide enough accuracy to be used for flame tracking in the
outer layers of WD. Next order in $1/\Lambda$ seems marginally sufficient
at $f_0\le 0.2$ yet computations become too involved in $r\ne 0$ case
(Sec.~4.4). More importantly, flame capturing as presented is a
general method, thus it is desirable to get results with larger range
of validity in a ready to use form. In this section we present fits
neatly interpolating between small and large $\Lambda$ regions and then
summarize the procedure for getting $K$ ($\tilde{K}$) and $R$ for \p{I1}
(respectively \p{artvisK}) in the SN Ia simulations.

For model I and its modification in Sec.~4.4
\begin{equation}\label{mf0}
d(\Lambda)=\frac{m_1}{1+m_2/\Lambda}+\frac{m_3}{(1+m_4/\Lambda)^2}
\end{equation}
with $m_{1\ldots 4}$ obtained at each $f_0$ to minimize the mean square
deviation (with weights proportional to actual $d(\Lambda)$)
was the simplest fit found.\footnote{%
Different fits must be used for model II, as the asymptotic behavior of $d$
as $\Lambda\to 0$ is different. The latter was not found analytically, but numerically
we observed that $d$ scaled as some power of $\Lambda$
($\exists\alpha\in(0;1):\,d/\Lambda^\alpha\to const$). $d(\Lambda)=\widetilde{m}_1
\bigl(1+\frac{\widetilde{m}_2}\Lambda+\widetilde{m}_4\Lambda^{-\widetilde{m}_5}\bigr)^{-\widetilde{m}_3}$
provided 0.3\% accuracy for all $f_0$ studied, and 0.45\% accurate fits were achieved
with the same expression when $\widetilde{m}_5=2$ was fixed. $\widetilde{m}_3\widetilde{m}_5$
varied within $[0.22;0.26]$, this serving as an estimate for the above $\alpha$. 
Fits in the region $1/\Lambda\in[10^5;10^9]$ provide strong evidence that $\forall f_0\: \alpha=0.25$.}
The values shown in the graph below guarantee 0.2\%
accuracy for each $f_0\in\{0.01;0.025(0.025)0.975\}$ and
$\Lambda\in[10^{-3};10^5]$ studied (for any $\Lambda$ in fact, as comparison
of asymptotes shows). For $f_0>0.3$ agreement is significantly better.

\begin{figure}[htbp] \begin{centering}
\includegraphics[width=0.6\textwidth]{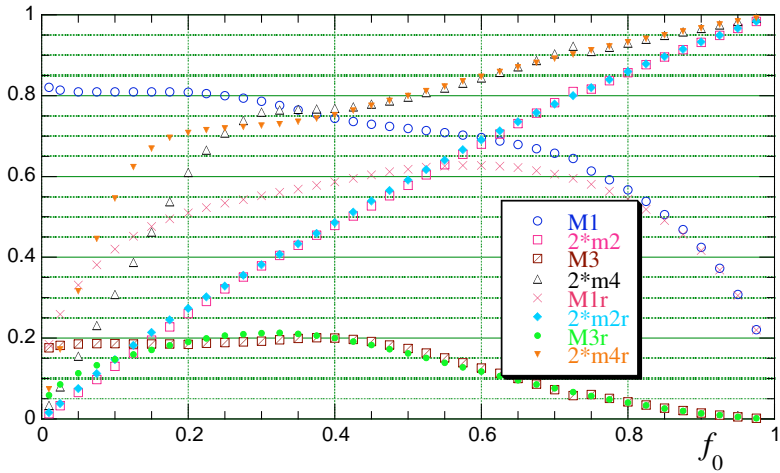}
\caption{Fit \p{mf0} parameters for $r=0$ and $r=3/4$ (the last 4 curves according to the legend).
$M_{1,3}=m_{1,3}\sqrt{f_0}$.}
\end{centering}\end{figure}

I have not succeeded to find a simple expression for $m_{1\ldots 4}(f_0)$
providing good agreement for all expansions; this is due to delicate
interplay between the two terms near $1/\Lambda=0$.\footnote{complicated form of
$m_{1\ldots 4}(f_0)$ is related to different significance of the two asymptotic
regions in the fit: for $f_0=0.01$ the $d(\Lambda)$ becomes reasonably linear
($d/\Lambda\approx const$) only at
$1/\Lambda>300$, whereas for $f_0>0.5$ this happens at $1/\Lambda\ge 2$.}
To overcome this a number of other possible fits were tested. In short,
3-parameter fits were significantly less accurate (though 3\% accuracy was
achieved, uniformly on $\Lambda\in(0;\infty)$), fits with more than 4 parameters did not yield significant
improvements; none of accurate fits admitted simple expressions for its
coefficients in terms of $f_0$.

This is not a major issue as $f_0$ is a parameter one can fix at some
convenient value throughout the simulation.
In \cite{X95} $f_0$ was fixed at 0.3, which essentially yielded thinnest
flames throughout the $\Lambda$ range in the problem\footnote{A. Khokhlov,
private communication. Note that this agrees with our findings, cf. Fig.~6.}.
The values for $K$ and $R$ used there result in
actual flame speed $D_f=d(f_0,\Lambda)\sqrt{f_0}S$, reasonably close to $S$ at
small $1/\Lambda$ and $ f_0=0.3$ (7\% smaller in the WD
center, however $\sim 1.45$ times smaller when $\rho=3\times 10^7$, 0.5C+0.5O), and flame width
$W\simeq w_2(f_0,\Lambda)\beta\Delta x$; $\beta=1.5$ was used, $\Delta x$ is a grid size.
Now, the prescription we advocate for normalizes
width as well as the flame speed, thus the logical criterion for $f_0$ to
suggest would be the fastest tail (near $f=0$) decay in the profile normalized
to unit width $w_2$ (we discuss traditional FCT realization \p{I1} here;
logic is the same as in 4.4).
This translates to smallest $1/w_2d$. This, however, can be made
arbitrarily small by taking sufficiently small $f_0$; I would suggest to
stick to $f_0=0.2$, as this yields $1/w_2d$ least sensitive to $\Lambda$ in its
interval of interest, from 0.4 to $\infty$; normalized this way flame profiles
also exhibit fairly low sensitivity to $\Lambda$. At smaller $f_0$ at small $\Lambda\,(\leq 0.5)$
profile shapes on $f\in [0.1;1]$ become rather nonlinear, making the choice of $w_2$
as a measure for \lq\lq width\rq\rq\ more questionable.
%\textcolor{red}{We will comment more
%on this in the next section, and will propose a model without tails whatsoever.}

One can observe that $\bigl(m_1+m_3\bigr)(f_0)$ is a significantly smoother
function than both terms separately (for any $r$). This sum obviously equals
$d_0\equiv d(f_0,r|\Lambda=\infty)$ with stated above accuracy ($\le 0.2$\%). The
latter is perfectly well-defined
(and is a solution of \p{Lambdainf} in the $r=0$ case). $d(f_0,0|\Lambda=\infty)$ can be approximated as \[
d_0=\left[\left(1-f_0^{1.4448}\exp(0.58058(1-f_0^{-1})\,)\right)\Bigm/ f_0
\right]^{0.5}, \]
with accuracy of 0.64\%; for $r=1/2$ and $r=3/4$
$d_0=\left[2\left(1-f_0\right) f_0^{r-1}\right]^{1/2}$ provides 1.5\% accurate fits.\\

%\flushleft {\Large \center \bf Andriy V.Zhyglo}\qquad \qquad {\sc \large addendum (awards)\ }\\ \bigskip
{\center \begin{tabular}{|r|r|rrrr|r|}\hline 
$f_0$ & $r\;$ & $m_1\;\;$ & $m_2\;\;$ & $m_3\;\;$ & $m_4\;\;$ & $\epsilon_d,\,10^{-4}\%$ \\ \hline
0.1& & 2.9248& 0.081016& 0.23690& 0.32153& 90\\ 
0.2& & 1.9588& 0.14574& 0.26929& 0.42824& 66\\
0.3& & 1.4620& 0.19389& 0.32605& 0.39759& 11\\
0.4& & 1.1889& 0.24193& 0.30495& 0.39044& 6.1\\
0.5& & 1.0312& 0.29358& 0.23117& 0.40545& 11\\
0.6& & 0.90646& 0.34328&0.15479& 0.42613& 7.2\\
0.9& & 0.44817& 0.46634&0.015034& 0.48381& 5.3\\ 
\hline
0.2&3/4& 1.1683& 0.14139& 0.40205& 0.37378& 56\\
0.6&3/4& 0.81846& 0.34834& 0.14334& 0.42779& 9.0\\ \hline
0.2&1/2& 1.3969& 0.14439& 0.35064& 0.39886& 52 \\ \hline

\end{tabular}%}
\\ \bigskip
Table 1: Parameters of fit \p{mf0}. The last column shows the maximum
relative discrepancy between the $d(\Lambda)$ and its fit 
(with $m_{1\ldots 4}$ truncated exactly as in the Table) at $\Lambda\in [1/3;20]$,
$\epsilon_d=(|\Delta d|/d)_\mathrm{max}$. The first seven entries correspond to
model \p{I1}, $r=0$ effectively.}\medskip

The fit parameters $m_{1\ldots 4}$ for $f_0=0.1\ldots 0.9$, as well as for the
$\lbrace f_0,r\rbrace$ values suggested in 4.4 optimized for
$\Lambda\in [1/3;50]$ are summarized in Table 1. The final strategy is to pick one\rq s favorite $f_0$
(or $\lbrace f_0,r\rbrace$) and corresponding $m_{1\ldots 4}$ from the table;
then at each computation step determine the local
$\Lambda$ \p{lamb} and \emph{needed} $D_f(\Delta)$ (based on local density, gravity, etc.);
\p{mf0} then defines $d$, and
\p{w2exp} (respectively \p{w312} or \p{w334}) determines dimensionless width
$w$ ($w_2$ or $w_3$). Then
\begin{equation}\label{KR}
K=\frac{D_f}d\cdotp\frac{W}w,\quad  R=\frac{D_f}d\Bigm/\frac{W}w,
\end{equation}
(where $W$ is a desired flame width, say 3 or 4 computational cells) will lead
to a flame with the desired speed $D_f$ and width $W$ within the approximation
adopted in this paper (see Appendix).

\section{Conclusions}
We studied a steady state 1D burning governed by a diffusion
equation for one species (fuel) with a source term representing fuel consumption;
the latter depended on fuel concentration $1-f(\in[0;1])$ only. A 3D
generalization of this system is widely used as a fiducial field governing the flame
position and heat release in non-steady 3D simulations. Finding correct
parameters of this generalization was the principal goal of the paper.

A numerical code was written for integrating the equation and finding the
flame velocity $D$ and width $W$. The problem was
reduced (by factoring out artificial diffusivity and burning rate normalization)
to finding dimensionless $d$ and $w$ depending on the burning rate shape $\Phi_0(f)$ and
Atwood number $\ma$ only. Several $\Phi_0$\rq s were studied to estimate
the accuracy of the technique. The results of \cite{KPP}
were generalized for the system with gas expansion, the spectrum of $d$ was
found continuous as in the original paper (spectrum dependence on the
$\Phi_0(f)$ behavior near 0 and 1 was investigated in general as well; KPP-like velocity spectrum and flame profiles were found characteristic of the burning rates decreasing linearly or faster at $f\to 0$ and 1); this was confirmed numerically. Analytic solution for two other burning rate -- equation of state combinations were obtained,
reducing the
parameters finding to solving a transcendental equation (in Bessel 
functions). Taking bulk integration step of 0.001 (smaller near
peculiarities) in direct integration (in $f$) was enough to ensure coincidence of the results with these
analytic ones within 1\% for both $d$ and $w$.%(these are the maximum
%deviations; mean quadratic discrepancy was 0.3\% for $d$, 0.25\% for $w$)

Asymptotic
behavior of $d$ and $w$ at $\ma$ near 1 (no expansion) and $\infty$ (large
expansion) was established. Interpolating formulae were proposed for
these quantities valid for any expansion coefficient, thus providing an
efficient method for finding the parameters for the equation describing
flame propagation in 3D simulations. Another variation of the scheme was
proposed and studied, which reslulted in finite flame width, hence promising better
performance in non-stationary simulations.
Specific realizations were pointed out, effectively uniformizing flame profiles
(further improvement beyond widths normalization, advocated for with any
variation of \p{I1}),
thus ensuring the same performance at all stages of simulation.

Brief analysis showed that the flame
shall propagate with the prescribed velocity $D_f$ and shall have the prescribed width $W$
with high accuracy, being
insensitive to details of the actual burning rate as long as most of the
energy is released within its width and as long as
the matter enthalpy scaling $H\propto 1/\rho$ holds on flame width scale.
This is the case in the SN Ia problem, one
usually defines a \lq\lq flame\rq\rq\ for the first condition to hold, and the second
one is true for the first stage of deflagration in a white dwarf (while all the
effectively coupled components of the matter can be considered as ultrarelativistic
gas), in which
virtually all of the nuclear energy is released.

We argued that the restriction of burning rate depending on $f$ only
is a good approximation; burning inside the flame in most simulations
is described by accurately modeling burning out of one element, while
effect of burning of other species estimated crudely. This is physical ---
if several reactions have comparable rates, they can be effectively
described by one $f$ variable; if one reaction is significantly faster,
the flame is influenced by it to a higher degree, and the approximate
estimate is enough.

For the boundary case, when different reactions may be important in
different flame zones (or when $H\rho$ changes significantly within the flame)
actual velocity of the artificial flame speed
may differ from the value based on which the FCT was modeled. One might
then try to consider steady state solutions of case-specific generalizations of \p{I1} with several
species and temperature involved. As another class of effects caused by
finite flame width are
effects of flame curvature and matter velocity gradients on the flame
propagation. These will be a subject of future publications, as will be
testing of the results.\\

\vspace*{0.25cm}%\indent\hspace{3mm}
\noindent{\Large \bf{Appendix}}\vspace{0.25cm}\\
%\end{center}
\nopagebreak[4]Here I present some expectations of the FCT performance with the new
parameters in real-world simulations. Basic setup is the following:
the physical system consists of some reactant, its concentration
denoted $g(\mathbf{r},t)$, $\mathbf{u},\rho$ denote its velocity and
density fields. The burning is modeled as described in Introduction.
First consider a steady state burning; in such the quantities are
distributed according to \p{artvisfront0} and (1D without loss of generality)
{\setlength\arraycolsep{2pt}
\begin{eqnarray}\label{fg}
%\begin{array}[lcr]\label{fgg}
\frac{dg}{dx}&=&\frac{G(H,g)}{\rho_0 D_f}\\ \label{fgh}
\frac{dH}{dx}&=&q\frac{df}{dx},%\label{fgh}
%\end{array}
\end{eqnarray}}%
continuity equation and equation of state having been used to eliminate
$u(x)$ and express burning rate of $g$ as a function of enthalpy $H$
instead of the temperature (considering the pressure constant $H$ and $g$
are the only thermodynamically independent quantities. Equation for $g$
effectively decouples because of \p{fgh} form, dictated by FCT. $g$ is generally a vector,
describing concentrations of all the species modelled. In some realizations instead of
\p{fg} one simply determines $g$ as some pre-defined function of $f$,
that making the analysis here more straightforward).

A crucial observation at this point is that $H=\frac{const}{\rho}$ would be
enough to get (from \p{fgh}) an expression for $u$ in \p{artvis0} coincident with that used in
model I of Sec.~2, thus tautologically yielding the same $D$ (and then the same $W$).
Rest of the equations would
not matter: given $f$ \p{fgh} would determine $H(x)$, while \p{fg} would define
$g(x)$. The boundary conditions for $H$ and $g$ are satisfied automatically:
gradient of $H$ tends to zero at infinities; that of $g$ behaves the same at
$+\infty$, assuming as usual fast reaction rate decay at low temperatures; and
$g(-\infty)=0$ for usually considered $G(H,g)\propto g^m,m\ge 2$ and
$G(H(-\infty),g(+\infty))$ not very small (otherwise no reaction proceeds). 
On the other hand, different $H(\rho)$ would yield different $u(x)$, and the
resulting eigenvalue for $D$ might differ drastically. As a ready example
consider model II of Sec.~2 for prescribing $K,R$. Suppose in reality $H$ does
scale $\propto 1/\rho$. One can check that the actual flame speed will be
$d_\mathrm{II}/ d_\mathrm{I}$ times smaller (that is 7.6 times smaller for
$f_0=0.9$,$\: 1/\Lambda=20$); one can expect comparable discrepancy for model I used
for modeling a system with $d\frac 1\rho/dH|_{p=const}$ being much smaller at $H(+\infty)$
(cold fuel) than at $H(-\infty)$ (hot ash). Thus, the distribution of density within
the flame is important, not just an integral jump.

Fortunately for the method, $H=\frac{4p}{\rho}$ for ideal (=non-interacting in
statistical physics sense, degeneracy and statistics do not matter) ultrarelativistic gas,
and the matter in WD core can reasonably be considered as such. It remains
ultrarelativistic during the burning, thus making assumptions behind model I
applicability fulfilled.%
\footnote{Even when $H\rho$ changes appreciably, the way
it changes may be predicted. Then one must find the $D_f$ as eigenvalue
of the system \p{artvisfront0} and \p{fgh} beforehand and use thus obtained $K$ and
$R$ in \p{I1} for FCT with model I. When expansion is small chances are
better that the details of $\rho$ changing may be neglected.}
 There are no clear preferences as to what
value for $f_0$ to use at this stage.

In non-stationary burning, \p{masterKR} (which would have yielded the
prescribed flame velocity, time-dependent in this case) cannot be derived
from \p{I1}, $\mathbf{u}$ will explicitly depend on coordinates and time. Far
from caustics though one may hope that dependence to be a small correction
to $|\mathbf{u}(f)|$ from the previous paragraph. Changing cross-section of
streamlines (due to flame curvature and stretch), pressure changes
in the gravitational field (and thus violation of $H\rho=const$ relation%
\footnote{It may be shown, in fact, that the latter effect is similar to the flame
curvature with radius a few times smaller than the stellar radius, thus
negligible in comparison with true curvature effects. This hydrostatic pressure
variation must not be confused with pressure gradients caused by burning, effects of
which are small as $O(D_f^2/c^2)$, $c$ being the local sound speed; the latter effects
were completely ignored in this paper. Finally, as by background tangential velocity gradients
the flame will be affected by background sound waves.})
are examples of this phenomenon; quasi-steady state burning is possible in
this case, and the flame speed of such is a likely real flame speed in the
coupled system of hydrodynamic equations and \p{I1}. Linearizing the
perturbed $\mathbf{u}$ near the flame position one might find the effect of
nonstationarity being possible to model with some equivalent flame with
curvature. We found the geometric effects not very pronounced, being $\leq 5\%$
for flames with radius of curvature 10 times their width and above.
However this whole discussion is superficial in that conditions far from the
flame may still have a strong effect on its speed, as the boundary
conditions are at infinities (however the models with finite width, like those studied in Sec.~4.4, must be more robust). This was seen in Section~3, similar
complications occur in the problems with flame curvature. Real
field-testing of the flame evolution is needed for the final verdict.\\

\vspace*{0.25cm}%\indent\hspace{3mm}
\noindent{\Large \bf{Acknowledgements}}\vspace{0.25cm}\\ \nopagebreak
%\section{Acknowledgements}
I am grateful to Alexei Khokhlov for proposing this problem,
for fruitful discussions, suggestions and encouragement
during the work. 
This work was supported by the Department of Energy under Grant No. B523820 to the
Center for Astrophysical Thermonuclear Flashes at the University of Chicago.

\end{document}